\documentclass[conference]{IEEEtran}
\IEEEoverridecommandlockouts

\usepackage{cite}
\usepackage{amsmath,amssymb,amsfonts}
\usepackage{bm}
\usepackage{algorithmic}
\usepackage{graphicx}
\usepackage{textcomp}
\usepackage{xcolor}
\usepackage{amsthm}

\usepackage{algorithm}
\usepackage[acronym,shortcuts]{glossaries}
\usepackage{hyperref}

\newacronym{B5G}{B5G}{beyond fifth-generation}
\newacronym{6G}{6G}{sixth-generation}
\newacronym{SM}{SM}{spatial modulation}
\newacronym{GSM}{GSM}{generalized spatial modulation}
\newacronym{AirComp}{AirComp}{over-the-air computation}
\newacronym{AP}{AP}{activation pattern}
\newacronym{AWGN}{AWGN}{additive white Gaussian noise}
\newacronym{BER}{BER}{bit error rate}
\newacronym{BS}{BS}{base station}
\newacronym{CSI}{CSI}{channel state information}
\newacronym{GaBP}{GaBP}{Gaussian belief propagation}
\newacronym{VGaBP}{VGaBP}{vector Gaussian belief propagation}
\newacronym{ICC}{ICC}{integrated communication and computing}
\newacronym{i.i.d.}{i.i.d.}{independent and identically distributed}
\newacronym{IM}{IM}{index modulation}
\newacronym{ISAC}{ISAC}{integrated sensing and communication}
\newacronym{LMMSE}{LMMSE}{linear minimum mean squared error}
\newacronym{ML}{ML}{maximum likelihood}
\newacronym{MAP}{MAP}{maximum a posteriori}
\newacronym{MIMO}{MIMO}{multiple-input multiple-output}
\newacronym{MMSE}{MMSE}{minimum mean squared error}
\newacronym{MP}{MP}{message passing}
\newacronym{MSE}{MSE}{mean squared error}
\newacronym{MUI}{MUI}{multi-user interference}
\newacronym{NMSE}{NMSE}{normalized mean squared error}
\newacronym{QAM}{QAM}{quadrature amplitude modulation}
\newacronym{QPSK}{QPSK}{quadrature phase shift keying}
\newacronym{BPSK}{BPSK}{binary phase shift keying}
\newacronym{SNR}{SNR}{signal-to-noise ratio}
\newacronym{ZF}{ZF}{zero forcing}
\newacronym{RF}{RF}{radio frequency}
\newacronym{TDD}{TDD}{time-division duplex}
\newacronym{SGA}{SGA}{scalar Gaussian approximation}
\newacronym{IC}{IC}{interference cancellation}
\newacronym{SIC}{SIC}{soft interference cancellation}
\newacronym{mu-MIMO}{mu-MIMO}{multiuser-MIMO}
\newacronym{DoF}{DoF}{degrees of freedom}

\begin{document}

\title{Integrated Communication and Computing \\with Index Modulation
\vspace{-1ex}}

\author{
\IEEEauthorblockN{Shreesal Shrestha, Kuranage Roche Rayan Ranasinghe, Hyeon Seok Rou, and Giuseppe Thadeu Freitas de Abreu}
\IEEEauthorblockA{\textit{School of Computer Science and Engineering, Constructor University, 28759 Bremen, Germany} \\
Emails: [shrshresth, kranasinghe, hrou, gabreu]@constructor.university}
\vspace{-5ex}
}

\maketitle

\begin{abstract}
We propose an \ac{ICC} architecture that repurposes the inactive antennas of a spatial \ac{IM} transmitter to perform \ac{AirComp}.
While \ac{IM} delivers excellent spectral efficiency by selectively activating a subset of transmit antennas, the unused antennas are conventionally completely deactivated, forgoing spatial \ac{DoF} that can be exploited without additional bandwidth.
By instead transmitting a pre-equalized, low-power computing stream over these idle antennas, the proposed architecture achieves simultaneous data transmission and computation without requiring orthogonal frequency or time resources.
Alignment is performed entirely at the transmitters, such that the pre-equalization occurs leveraging local \ac{CSI}, while at the receiver, a \ac{VGaBP} detector recovers the data payload under the discrete \ac{IM} codebook constraint before the target function is estimated from the residual.
Numerical results against exact \ac{ML} baselines confirm near-optimal detection under the adopted statistical model, and reveal that the number of antennas assigned to data transmission governs a direct trade-off between modulation robustness and \ac{AirComp} accuracy.
\end{abstract}

\begin{IEEEkeywords}
ICC, AirComp, IM, message passing, GaBP.
\vspace{-1ex}
\end{IEEEkeywords}

\glsresetall

\vspace{-1ex}
\section{Introduction}
\vspace{-1ex}
The evolution of \ac{B5G} and \ac{6G} \cite{Saad_2020} wireless networks demands unprecedented spectral efficiency along with seamless integration of edge functionality \cite{Larsson_2014,Tataria_2021}.
Among the physical-layer solutions, \ac{IM} \cite{Mao_2019,Shamasundar_2022, Basar_2016, Rou_Asilomar2024, Rou_TWC_IM2025} has emerged as an appealing paradigm that strikes a highly favorable trade-off between spectral and energy efficiency.
By conveying information through the selective activation of a small subset of available resources, such as transmit antennas, spatial \ac{IM} or \ac{SM}/\ac{GSM} \cite{Mesleh_2006, Younis_2010, Rou_QSPM_TWC2022} enhances spectral efficiency over single-antenna systems while reducing the number of active \ac{RF} chains and their associated power consumption.

Currently, the push toward shared spectrum and hardware has accelerated research into multifunctional wireless systems.
\Ac{ISAC} \cite{Wild_2021, Rou_TWC_ISAC2024, Zhang_2026,
Rayan_TWC_FIMISAC2026} has, to an extent, adopted \ac{IM} to embed communication bits into radar waveforms and to exploit sparse antenna-index modulation for simultaneous communication and target detection \cite{Elbir_2024}.
In turn, the realization of \ac{ICC} through \ac{AirComp} \cite{Nazer_2007, Ando_CAMSAP2023, Wang_2024, Ando_ITJ2025, Rou_ICASSP2026} has so far been pursued through general receiver frameworks \cite{Ranasinghe_2026, Du_2024}, leaving its integration with \ac{IM} largely unexplored.
Moreover, in conventional \ac{IM}, the unselected antennas are deactivated to minimize power and \ac{RF} hardware costs, such that, to the best of our knowledge, their use as a dedicated resource for \ac{AirComp} has not been considered.

In this paper, we propose an \ac{ICC} system that performs \ac{AirComp} over the inactive antennas of an \ac{IM} system, deliberately exchanging the hardware savings of \ac{IM} for edge computing capabilities.
By fully activating the transmitter array, each user dedicates some antennas to \ac{IM} data transmission while radiating a pre-equalized, low-power computing stream over the remaining inactive set.
Since the two components occupy disjoint antenna sets, the transmit power splits exactly between them and the computing stream can be aligned from local \ac{CSI} without disturbing the data payload.
The sparsity of the \ac{IM} waveform is in fact what creates a spatial subspace that is idle by construction, such that each \ac{RF} chain radiates either a constellation symbol or a computing sample, and never their superposition.
Operating the computing stream at a fraction of the data power further ensures it superimposes over the wireless channel as background interference, preserving detectability of the primary data.

Recovering the user payloads at the \ac{BS}, however, remains challenging, as exact \ac{ML} detection requires an exhaustive search over all combinations of activation patterns and data symbols, whose cost grows exponentially with the number of users.
We therefore adopt an efficient \ac{VGaBP} detector \cite{Rou_2022}, retaining exact \ac{ML} in small-scale scenarios to establish performance bounds.
Our results show that the proposed transmitter-side alignment outperforms a channel-blind transmitter under equal radiated computing power, and reveal a spatial design trade-off in which allocating more antennas to data transmission improves modulation robustness at the expense of coherent alignment.

\vspace{-1ex}
\section{System Model}
\label{sec:sysmodel}
\vspace{-1ex}
\subsection{Uplink ICC Signal Model}
\vspace{-1ex}
We consider an uplink \ac{MIMO} system consisting of $U$ users, indexed by $u \in \mathcal{U} \triangleq \{1, \dots, U\}$.
Each user is equipped with $M$ transmit antennas and transmits concurrently to a single \ac{BS} equipped with $N$ receive antennas.
Under perfect symbol and carrier phase synchronization among users, the received signal at the \ac{BS} at a given channel use is given by
\vspace{-1ex}
\begin{equation}
\vspace{-1ex}
\bm{y} = \sum_{u=1}^{U}\bm{H}_u\bm{x}_u + \bm{n} \in \mathbb{C}^{N},
\label{eqn:rx}
\end{equation}
where $\bm{x}_u \in \mathbb{C}^{M\times 1}$ is the transmit signal of user $u$, conveying both communication and computing information, subject to the total power constraint $\mathbb{E}[\|\bm{x}_u\|^2] \leq P_{\max}$, and $\bm{n} \sim \mathcal{CN}(\bm{0}, \sigma^2\bm{I}_N)$ is the \ac{AWGN} vector with variance $\sigma^2$.

The channel $\bm{H}_u \in \mathbb{C}^{N \times M}$ follows the uncorrelated block Rayleigh fading model typical in the \ac{AirComp} literature \cite{Nazer_2007, Ando_CAMSAP2023, Ando_ITJ2025, Wang_2024}, with \ac{i.i.d.} entries $[\bm{H}_u]_{n,m} \sim \mathcal{CN}(0,1)$ constant over a fading block.
We assume that the \ac{BS} has perfect knowledge of all channels, while each user $u$ is restricted to local \ac{CSI} $\bm{H}_u$.
The latter constitutes the only \ac{CSI} required at the transmitters, such that no inter-user \ac{CSI} exchange is necessary beyond the scalar $\eta$, broadcast once per fading block, introduced in Section~\ref{sec:computing}.

\vspace{-1ex}
\subsection{Index-Modulated Transmit Signal}
\vspace{-1ex}
\label{sec:im}
At every channel use, each user employs only $k<M$ of its transmit antennas for data transmission, reserving the remaining $M-k$ for the computing signal.
The specific set of active data antennas for user $u$ is denoted by the activation pattern $\mathcal{A}_u$. 
To allow binary encoding, $\mathcal{A}_u$ is drawn from a predefined codebook 
\vspace{-1ex}
\begin{equation}
\vspace{-1ex}
\mathcal{Q} \triangleq \{\mathcal{S}_1, \dots, \mathcal{S}_Q\},
\label{eqn:apset}
\end{equation}
comprising $Q \triangleq |\mathcal{Q}| = 2^{\lfloor \log_2 \binom{M}{k} \rfloor}$ admissible patterns, where each $\mathcal{S}_q \subset \{1, \dots, M\}$ and $|\mathcal{S}_q| = k$.

The codebook is constructed to be balanced such that every antenna appears in exactly $kQ/M$ patterns%
\footnote{The balance condition requires only that every antenna appear in $kQ/M$ patterns, which any such balanced subset of size $Q$ satisfies. For the configurations considered here, one is obtained by cyclically shifting a suitable starting pattern over the $M$ antennas, yielding $M$ patterns in which every antenna appears exactly $k$ times.}
ensuring a uniform activation probability of $\Pr[m \in \mathcal{A}_u] = k/M$ \cite{Wei_2019}.
User $u$ maps $\log_2 Q$ index bits to select a pattern $\mathcal{A}_u \in \mathcal{Q}$, and $k \log_2 |\mathcal{X}|$ symbol bits onto $k$ data symbols $\{v_{u,m}\}_{m \in \mathcal{A}_u}$, where $v_{u,m} = \sqrt{E_v}d_{u,m}$ and $d_{u,m}$ is drawn from a unit-energy $|\mathcal{X}|$-ary constellation $\mathcal{X} \subset \mathbb{C}$ with $\mathbb{E}[|d_{u,m}|^2] =1$, with $E_v$ denoting the per-symbol data energy.
Each symbol $v_{u,m}$ is radiated over its corresponding active antenna $m$, such that the $k$-sparse data vector is
\vspace{-1ex}
\begin{equation}
\vspace{-1ex}
\bm{v}_u \triangleq \sum_{m \in \mathcal{A}_u} v_{u,m}\bm{e}_m \in \mathbb{C}^{M \times 1},
\quad
\|\bm{v}_u\|_0 = k, \text{ for } 0 \notin \mathcal{X}
\label{eqn:datavec}
\end{equation}
where $\bm{e}_m$ denotes the $m$-th column of $\bm{I}_M$.

Since information is conveyed both by the transmitted symbols and by the choice of $\mathcal{A}_u$, the achievable rate per user (in bits per channel use) is
\vspace{-1ex}
\begin{equation}
\vspace{-1ex}
B = \underbrace{\big\lfloor \log_2 \tbinom{M}{k} \big\rfloor}_{\text{index bits}}
  + \underbrace{k \log_2 |\mathcal{X}|}_{\text{symbol bits}}.
\label{eqn:rate}
\end{equation}

\subsection{The AirComp Operation}
\label{sec:aircomp}
\ac{AirComp} exploits the superposition property of the wireless multiple-access channel to compute a function of the users' data directly over the air, enabling the \ac{BS} to recover the aggregate result rather than decoding individual measurements.
The supported class of functions are the nomographic functions \cite{Wang_2024, Liu_2020}, which admit the representation
\vspace{-1ex}
\begin{equation}
\vspace{-1ex}
f(s_1,\dots,s_U) = \phi\!\left(\sum_{u=1}^{U}\rho(s_u)\right),
\end{equation}
where $s_u$ is the local measurement of user $u$, and $\rho(\cdot)$ and $\phi(\cdot)$ are the pre- and post-processing functions applied at the user and at the \ac{BS}, respectively.
The pre-processed computing symbols $c_u \triangleq \rho(s_u)$ are modelled as \ac{i.i.d.} $\mathcal{CN}(0, \sigma_c^2)$ with $\sigma_c^2$ known at the \ac{BS} \cite{Ranasinghe_2026}.
For ease of exposition, we set both maps to the identity, such that the target function reduces to the arithmetic sum $f(\bm{c}) = \sum_u c_u$, any other nomographic function follows from an appropriate choice of the pre- and post-processing pair.

The computing symbol is distributed across the $M-k$ antennas left unoccupied by the \ac{IM} pattern.
To this end, letting $\bar{\mathcal{A}}_u \triangleq \{1,\dots,M\} \setminus \mathcal{A}_u$ denote the complement of the active antenna pattern, we define
\vspace{-1ex}
\begin{equation}
\vspace{-1ex}
\bm{E}_u \triangleq [\bm{e}_m]_{m \in \bar{\mathcal{A}}_u} \in \{0,1\}^{M \times (M-k)}
\end{equation}
to collect the columns of $\bm{I}_M$ indexed by $\bar{\mathcal{A}}_u$.%
\footnote{The computing stream need not occupy all of $\bar{\mathcal{A}}_u$. For large $M$, a far smaller subset suffices, but selecting it couples the choice to the codebook design, since a poor pairing may leave antennas unused or yield ill-conditioned computing channels. Such a joint design is left to future work.}

The transmit computing vector is then $\bm{E}_u\bm{w}_u c_u$, where the local precoding vector $\bm{w}_u \in \mathbb{C}^{(M-k) \times 1}$ applies the spatial alignment and is inherently data-dependent. Thus, it must be recomputed at every channel use.
This transmission is subject to the per-user computing power budget
\vspace{-1ex}
\begin{equation}
\vspace{-1ex}
\mathbb{E}_{\mathcal{A}_u, c_u}\left[\|\bm{E}_u\bm{w}_u c_u\|^2\right] = \sigma^2_c\mathbb{E}_{\mathcal{A}_u}\left[\|\bm{w}_u\|^2\right] \leq E_c,
\label{eqn:computing_budget}
\end{equation}
where the expectation is taken over $c_u$ and the activation pattern $\mathcal{A}_u$ for a given channel realization.
The specific design of $\bm{w}_u$ is detailed in Section~\ref{sec:computing}.

\vspace{-1ex}
\subsection{Transmit Signal and Power Allocation}
\vspace{-1ex}
\label{sec:txsignal}
The combined transmit signal of user $u$ is given by
\vspace{-1ex}
\begin{equation}
\vspace{-1ex}
\bm{x}_u \triangleq \underbrace{\bm{v}_u}_{\text{communication}} + \underbrace{\bm{E}_u\bm{w}_u c_u}_{\text{computing}} \in \mathbb{C}^{M \times 1}.
\label{eqn:tx}
\end{equation}

Since the two components are supported on the mutually exclusive spatial sets $\mathcal{A}_u$ and $\bar{\mathcal{A}}_u$, they admit the strict power decomposition $\|\bm{x}_u\|^2 = \|\bm{v}_u\|^2 + \|\bm{w}_u c_u\|^2$.
Taking the expectation over both symbols and the activation pattern, and invoking \eqref{eqn:computing_budget}, the average per-user transmit power is bounded by $\mathbb{E}[\|\bm{x}_u\|^2] \leq kE_v + E_c$, where $\mathbb{E}[\| \bm{v}_u\|^2] = kE_v$ holds for any realized pattern.
The system-wide power constraint is therefore satisfied provided
\vspace{-1ex}
\begin{equation}
\vspace{-1ex}
k E_v + E_c \leq P_{\max}.
\end{equation}

While this spatial split allows the computing signal to be pre-equalized independently without distorting the user's data, the wireless channel inevitably mixes the two subspaces, causing both components from all users to superimpose at the $N$-antenna receiver.

\vspace{-1ex}
\section{Precoding Design for AirComp}
\label{sec:computing}
\vspace{-1ex}

To recover the \ac{AirComp} function, the \ac{BS} projects the received signal onto a scalar using a fixed, publicly known, and unit-norm passive combining vector
\footnote{We remark that, under the \ac{i.i.d.} fading model considered, no fixed unit-norm vector is preferable to another, since $\bm{H}_u^{\mathrm{H}}\bm{a} \sim \mathcal{CN}(\bm{0}, \bm{I}_M)$ for any such $\bm{a}$, such that the specific form \eqref{eqn:passive_combiner} is adopted for concreteness.}
\vspace{-1ex}
\begin{equation}
\vspace{-1ex}
\bm{a} = \frac{1}{\sqrt{N}}\bm{1}_N. 
\label{eqn:passive_combiner}
\end{equation}

Since each user computes its local precoder from the effective channel seen through $\bm{a}$, the combining vector must be known at all transmitters.
A channel-adaptive combiner, by contrast, depends jointly on all user channels and must therefore be optimized at the \ac{BS} once per fading block and fed back to the transmitters \cite{Chen_2018, Ando_ITJ2025, Ando_CAMSAP2023}, replacing the scalar $\eta$ by $N$ complex coefficients per block.
As the focus of this work is the transmitter-side design, we adopt the fixed choice \eqref{eqn:passive_combiner}, at the cost of the receive array gain an optimized combiner would provide \cite{Chen_2018}.
Joint optimization of the combiner and the pattern-dependent precoder is left to future work.

Projecting the computing component of user $u$ onto $\bm{a}$ yields the effective computing channel
\vspace{-1ex}
\begin{equation}
\vspace{-1ex}
\bm{g}_u \triangleq \bm{E}_u^{\mathrm{H}} \bm{H}_u^{\mathrm{H}} \bm{a} \in \mathbb{C}^{(M-k)\times1},
\label{eqn:comp_channel}
\end{equation}
which varies at every channel use due to its dependence on the data-driven activation pattern.

To coherently align the target sum $\sum_u c_u$ at the \ac{BS}, the local precoder must satisfy the constraint
\vspace{-1ex}
\begin{equation}
\vspace{-1ex}
\bm{g}_u^{\mathrm{H}}\bm{w}_u = \eta,
\label{eqn:comp_constraint}
\end{equation}
where $\eta > 0$ is a network-wide target amplitude.

The corresponding minimum-norm \ac{ZF} solution is therefore
\vspace{-1ex}
\begin{equation}
\vspace{-1ex}
\bm{w}_{\text{zf},u} = \eta \frac{\bm{g}_u}{\|\bm{g}_u\|^2}.
\label{eqn:tx_zf}
\end{equation}

Notice that, while $\bm{E}_u$ leaves the radiated power unchanged, it determines which $M-k$ entries of $\bm{H}_u^{\mathrm{H}}\bm{a}$ are inverted by \eqref{eqn:tx_zf}, such that the power required for alignment is itself pattern-dependent.
Substituting the \ac{ZF} precoder \eqref{eqn:tx_zf} into the computing power budget \eqref{eqn:computing_budget} therefore yields the per-user feasibility condition
\vspace{-1ex}
\begin{equation}
\vspace{-1ex}
\eta \leq \eta_u \triangleq \frac{\sqrt{E_c}}{\sigma_c}\left( \frac{1}{Q} \sum_{q = 1}^Q \frac{1}{\| [\bm{H}_u^{\mathrm{H}}\bm{a}]_{\bar{\mathcal{S}}_q}\|^2}\right)^{-1/2}\!\!\!\!,\! \quad \forall u \in \mathcal{U},
\label{eqn:eta_feasibility}
\end{equation}
where $\bar{\mathcal{S}}_q \triangleq \{1,\dots,M\} \setminus \mathcal{S}_q$ denotes the inactive antenna set associated with the $q$-th pattern.

The \ac{BS} knows the channels $\{\bm{H}_u\}$ but not the realized patterns $\{\mathcal{A}_u\}$, and therefore averages the required precoder power over the equiprobable patterns of the codebook to obtain $\eta_u$. 
For coherent alignment common across users, it sets
\vspace{-1ex}
\begin{equation}
\vspace{-1ex}
\eta = \min_{u \in \mathcal{U}} \, \eta_u.
\label{eqn:eta_design}
\end{equation}

The \ac{BS} broadcasts \eqref{eqn:eta_design}, from which each user computes its local precoder \eqref{eqn:tx_zf} independently, once per fading block.

Two consequences of this design are worth noting.
First, since $\bm{a}$ is fixed and independent of the channel, the statistics of $\bm{H}_u^{\mathrm{H}}\bm{a}$ and therefore of $\eta$ do not scale with $N$, such that the computing stream obtains no receive array gain.
Second, while \eqref{eqn:eta_feasibility} ensures that each user exhausts its computing budget on average over the codebook, tying $\eta$ to the weakest user leaves the remaining users radiating below $E_c$, so that the reported \ac{NMSE} remains conservative.

\vspace{-1ex}
\section{Receiver Design}
\label{sec:receiver}
\vspace{-1ex}

Substituting \eqref{eqn:tx} into the uplink model, the received signal naturally separates into a data and a computing term as
\vspace{-1ex}
\begin{equation}
\vspace{-1ex}
\bm{y} = \sum_{u=1}^{U}\bm{H}_u\bm{v}_u + \underbrace{\sum_{u=1}^{U}\bm{H}_u\bm{E}_u\bm{w}_u c_u + \bm{n}}_{\tilde{\bm{n}}},
\label{eqn:obs}
\end{equation}
where, under the regime $E_c \ll E_v$, the computing signal is sufficiently weak to be treated as an effective noise vector $\tilde{\bm{n}}$ during detection, particularly since the individual symbols $c_u$ hold no relevance for data decoding.

Since each $\bm{v}_u$ is $k$-sparse, the receiver must resolve only $kU$ non-zero entries among the $MU$ transmitted, from $N$ receive dimensions.
However, the non-zero entries are not independent, being jointly constrained to the finite codebook \cite{Rou_2022}, given by
\vspace{-1ex}
\begin{equation}
\vspace{-1ex}
\mathcal{V} \triangleq \Big\{\sqrt{E_v}\sum_{m\in\mathcal{A}}d_m\bm{e}_m \;\Big|\; \mathcal{A}\in\mathcal{Q},\; d_m\in\mathcal{X}\Big\},
\label{eqn:validsetV}
\end{equation}
with $|\mathcal{V}| = Q|\mathcal{X}|^{k}$.

A standard scalar detector treating the entries independently cannot enforce this structural constraint, thereby allowing invalid patterns outside $\mathcal{Q}$ and destroying the index information.
We therefore adopt the vector-valued \ac{GaBP} detection framework of \cite{Rou_2022}, termed \ac{VGaBP} throughout, which assigns a single variable node to each user and thereby reduces the search from the $|\mathcal{V}|^U$ joint combinations of exhaustive \ac{ML} detection to $|\mathcal{V}|$ candidates per user, further reduced in \cite{Rou_TWC_IM2025}.
Conditioned on the channels and activation patterns, $\tilde{\bm{n}}$ is a linear combination of independent Gaussian variables and is therefore exactly Gaussian, with covariance
\vspace{-1ex}
\begin{equation}
\vspace{-1ex}
    \bm{\Sigma}_{\tilde{n}} \triangleq \sigma_c^2\sum_{u=1}^U\bm{b}_u\bm{b}_u^{\mathrm{H}} + \sigma^2\bm{I}_N,
    \label{eqn:Rexact}
\end{equation}
where $\bm{b}_u \triangleq \bm{H}_u\bm{E}_u\bm{w}_u$.

The \ac{BS}, however, cannot evaluate $\bm{b}_u$ as $\bm{E}_u$ encodes the unknown data payload, such that the effective noise it sees is instead a mixture of Gaussians, one per pattern combination. 
Knowing the $Q$ values that $\bm{b}_u$ may take, the \ac{BS} replaces each user's contribution by its expectation over them, modeling the mixture by a single Gaussian of matched second-order moment, which under the \ac{SGA} is further approximated by its diagonal, giving
\begin{equation}
\bm{\sigma}^2_{\tilde{n}} = \mathrm{diag}\Big( \sigma_c^2\sum_{u=1}^{U}\mathbb{E}_{\mathcal{A}_u}\big[\bm{b}_u\bm{b}_u^{\mathrm{H}}\big] \Big) + \sigma^2\bm{1}_N \in \mathbb{R}^{N\times1}.
\label{eqn:Rexpected}
\end{equation}

\subsection{Vector-Valued GaBP Detection}
\label{sec:gabp}
The rules below follow \cite{Rou_2022} with the quantities complex-valued rather than IQ-decoupled, and with a single variable node per user whose support is the codebook $\mathcal{V}$.

\textbf{Soft Interference Cancellation:} Subtracting the interference of all other users from the observation at the $n$-th receive antenna isolates the signal of user $u$, with residual variance $\psi_{n,u}$, as
\vspace{-1ex}
\begin{equation}
\vspace{-1ex}
\tilde{y}_{n,u} = y_n - \sum_{u'\neq u}\bm{h}_{n,u'}^{\mathrm{H}}
\hat{\bm{v}}_{n,u'} = \bm{h}_{n,u}^{\mathrm{H}}\bm{v}_u + \varepsilon_{n,u},
\label{eqn:softic}
\end{equation}
\vspace{-0.5ex}
\begin{equation}
\psi_{n,u} = \sum_{u'\neq u}\bm{h}_{n,u'}^{\mathrm{H}}\bm{\Psi}_{n,u'}
\bm{h}_{n,u'} + \sigma^2_{\tilde{n},n},
\label{eqn:condvar}
\end{equation}
where $\bm{h}_{n,u} \triangleq ([\bm{H}_u]_{n,:})^{\mathrm{H}} \in \mathbb{C}^{M\times1}$ is the local channel vector, $\hat{\bm{v}}_{n,u}$ and $\bm{\Psi}_{n,u}$ denote the soft replica of $\bm{v}_u$ and its error covariance, $\sigma^2_{\tilde{n},n}$ is the $n$-th element of \eqref{eqn:Rexpected}, and $\varepsilon_{n,u}$ is approximated as Gaussian under the \ac{SGA}.

\textbf{Belief Generation:} Excluding the $n$-th incoming message to prevent self-interference, the extrinsic belief is parameterized by the information vector $\bm{\mu}_{n,u}$ and precision matrix $\bm{\Lambda}_{n,u}$, and evaluated for a candidate $\bm{v} \in \mathcal{V}$ as
\vspace{-1ex}
\begin{equation}
\vspace{-1ex}
\ell_{n,u}(\bm{v}) \propto \exp \!\left( 2\Re\{\bm{\mu}_{n,u}^{\mathrm{H}}\bm{v}\} - \bm{v}^{\mathrm{H}}\bm{\Lambda}_{n,u}\bm{v} \right)
\label{eqn:extrinsic_belief}
\end{equation}
where
\vspace{-1ex}
\begin{equation}
\vspace{-1ex}
\bm{\mu}_{n,u} = \sum_{n'\neq n} \frac{\tilde{y}_{n',u}}{\psi_{n',u}} \bm{h}_{n',u},
\quad
\bm{\Lambda}_{n,u} = \sum_{n'\neq n} \frac{\bm{h}_{n',u} \bm{h}_{n',u}^{\mathrm{H}}}{\psi_{n',u}}.
\label{eqn:belief}
\end{equation}

\textbf{Soft Replica Generation:} Since the data prior is uniform across $\mathcal{V}$, it cancels during normalization, such that the Bayes-optimal soft replica and its error covariance follow directly from \eqref{eqn:extrinsic_belief} as
\vspace{-1ex}
\begin{subequations}
\begin{align}
\vspace{-1ex}
\hat{\bm{v}}_{n,u} &= \frac{\sum_{\bm{v}\in\mathcal{V}}\bm{v}\,\ell_{n,u}(\bm{v})}
{\sum_{\bm{v}\in\mathcal{V}}\ell_{n,u}(\bm{v})},
\label{eqn:denoiser_a}\\[2pt]
\bm{\Psi}_{n,u} &= \frac{\sum_{\bm{v}\in\mathcal{V}}\bm{v}\bm{v}^{\mathrm{H}}\ell_{n,u}(\bm{v})}
{\sum_{\bm{v}\in\mathcal{V}}\ell_{n,u}(\bm{v})}
- \hat{\bm{v}}_{n,u}\hat{\bm{v}}_{n,u}^{\mathrm{H}}.
\label{eqn:denoiser_b}
\end{align}
\end{subequations}

To prevent premature convergence during message passing, both variables are subsequently damped across iterations by a factor $\beta \in (0,1]$.

\textbf{Consensus:} Upon termination, the beliefs of all factor nodes are fused via \eqref{eqn:belief} with the summations extended over all $n$, yielding $\ell_u(\bm{v})$, from which the soft posterior mean and uncertainty covariance follow as
\vspace{-1ex}
\begin{subequations}
\begin{align}
\vspace{-1ex}
\bar{\bm{v}}_u &=
\frac{\sum_{\bm{v}\in\mathcal{V}}\bm{v}\,\ell_u(\bm{v})}
{\sum_{\bm{v}\in\mathcal{V}}\ell_u(\bm{v})},
\label{eqn:soft_mean}\\[2pt]
\bar{\bm{\Psi}}_u &=
\frac{\sum_{\bm{v}\in\mathcal{V}}\bm{v}\bm{v}^{\mathrm{H}}\ell_u(\bm{v})}
{\sum_{\bm{v}\in\mathcal{V}}\ell_u(\bm{v})}
- \bar{\bm{v}}_u\bar{\bm{v}}_u^{\mathrm{H}}.
\label{eqn:soft_cov}
\end{align}
\end{subequations}
The hard estimate for data detection is then obtained via the \ac{MAP} rule as
\vspace{-1ex}
\begin{equation}
\vspace{-1ex}
\tilde{\bm{v}}_u = \arg\max_{\bm{v} \in \mathcal{V}} \, \ell_u(\bm{v}),
\label{eqn:hard}
\end{equation}
such that the symbol bits are recovered from the non-zero entries of $\tilde{\bm{v}}_u$ and the index bits from its support $\hat{\mathcal{A}}_u = \mathrm{supp}(\tilde{\bm{v}}_u)$. 
To enable accurate interference cancellation, the soft statistics $\bar{\bm{v}}_u$ and $\bar{\bm{\Psi}}_u$ are forwarded to the subsequent \ac{AirComp} receiver.

\vspace{-1ex}
\subsection{AirComp Function Estimation}
\label{sec:func_estimation}
\vspace{-1ex}
Subtracting the expected data contribution from the received signal isolates the \ac{AirComp} transmission:
\vspace{-1ex}
\begin{equation}
\vspace{-1ex}
\bm{y}_{\mathrm{res}} \triangleq \bm{y} -\sum_{u=1}^U\bm{H}_u\bar{\bm{v}}_u = \sum_{u=1}^{U}\bm{b}_u c_u
+ \sum_{u=1}^{U}\bm{H}_u\check{\bm{v}}_u + \bm{n},
\label{eqn:sic}
\end{equation}
where $\check{\bm{v}}_u \triangleq \bm{v}_u - \bar{\bm{v}}_u$ is the soft data detection error. 

Applying the passive combiner $\bm{a}$ projects this residual vector into a scalar observation
\vspace{-1ex}
\begin{equation}
\vspace{-1ex}
r = \bm{a}^{\mathrm{H}}\bm{y}_{\mathrm{res}}
= \sum_{u=1}^{U}\bm{g}_u^{\mathrm{H}}\bm{w}_u c_u
+ \bm{a}^{\mathrm{H}}\Big(\sum_{u=1}^{U}\bm{H}_u\check{\bm{v}}_u\Big)
+ {n}_a,
\label{eqn:readout}
\end{equation}
where ${n}_a \triangleq \bm{a}^\mathrm{H}\bm{n} \sim \mathcal{CN}(0,\sigma^2)$ is the projected thermal noise.

Since the precoder enforces $\bm{g}_u^{\mathrm{H}}\bm{w}_u = \eta$ for every realized activation pattern, the target sum $\sum_u c_u$ aligns at amplitude $\eta$, corrupted only by the residual $\check{\bm{v}}_u$ and thermal noise, such that the computing stream is affected by detection errors only through an uncertainty that $\bar{\bm{\Psi}}_u$ accounts for.
Leveraging the uncertainty, the covariance of the aggregate \ac{SIC} error is approximated as
\vspace{-1ex}
\begin{equation}
\vspace{-1ex}
\bm{\Sigma}_{\mathrm{SIC}} = \sum_{u=1}^{U}\bm{H}_u\bar{\bm{\Psi}}_u\bm{H}_u^{\mathrm{H}}.
\label{eqn:sigmasic}
\end{equation}

\vspace{-2ex}
\begin{algorithm}[H]
\caption{IM-based GaBP Data Detection and AirComp}
\label{alg:gabp}
\begin{algorithmic}[1]
\STATE \textbf{Input:} $\bm{y}$, $\bm{H}_u$, $\sigma_{\tilde{n},n}^2$, $\eta$
\STATE \textbf{Parameters:} $\tau_{\max}$, $\beta \in (0,1]$
\STATE \textbf{Initialization:}
\STATE Set $\hat{\bm{v}}_{n,u}^{(0)} = \bm{0}$, $\bm{\Psi}_{n,u}^{(0)} = \frac{kE_v}{M} \bm{I}_M$ and $\tau = 0$
\REPEAT
    \STATE $\tau \leftarrow \tau + 1$
    \STATE Update soft IC residual $\tilde{y}_{n,u}$ via \eqref{eqn:softic}
    \STATE Update residual variance $\psi_{n,u}$ via \eqref{eqn:condvar}
    \STATE Compute beliefs $\bm{\mu}_{n,u}$ and $\bm{\Lambda}_{n,u}$ via \eqref{eqn:belief}
    \STATE Generate soft replica $\hat{\bm{v}}_{n,u}^{(\tau)}$ via \eqref{eqn:denoiser_a}
    \STATE Generate error covariance $\bm{\Psi}_{n,u}^{(\tau)}$ via \eqref{eqn:denoiser_b}
    \STATE Damp replica as $\hat{\bm{v}}_{n,u}^{(\tau)} \leftarrow \beta\hat{\bm{v}}_{n,u}^{(\tau)} + (1-\beta)\hat{\bm{v}}_{n,u}^{(\tau-1)}$
    \STATE Damp covariance as $\bm{\Psi}_{n,u}^{(\tau)} \leftarrow \beta\bm{\Psi}_{n,u}^{(\tau)} + (1-\beta)\bm{\Psi}_{n,u}^{(\tau-1)}$
\UNTIL{$\tau = \tau_{\max}$}
\STATE \textbf{Data Detection:}
\STATE Compute soft posterior mean $\bar{\bm{v}}_u$ via \eqref{eqn:soft_mean}
\STATE Compute soft posterior covariance $\bar{\bm{\Psi}}_u$ via \eqref{eqn:soft_cov}
\STATE Obtain hard estimates $\tilde{\bm{v}}_u$ via \eqref{eqn:hard}
\STATE \textbf{AirComp Function Estimation:}
\STATE Isolate AirComp residual $\bm{y}_{\mathrm{res}}$ via \eqref{eqn:sic}
\STATE Project to scalar observation $r$ via \eqref{eqn:readout}
\STATE Compute $\bm{\Sigma}_{\mathrm{SIC}}$ via \eqref{eqn:sigmasic} and $\sigma^2_{\mathrm{tot}} = \bm{a}^{\mathrm{H}}\bm{\Sigma}_{\mathrm{SIC}}\bm{a} + \sigma^2$
\STATE Calculate LMMSE weight $\zeta_{\mathrm{lmmse}}$ via \eqref{eqn:wlmmse}
\STATE Evaluate target function $\hat{f} = \zeta_{\mathrm{lmmse}} r$
\end{algorithmic}
\end{algorithm}
\vspace{-2ex}

With $\sigma_{\mathrm{tot}}^2 = \bm{a}^{\mathrm{H}}\bm{\Sigma}_{\mathrm{SIC}}\bm{a} + \sigma^2$ the scalar \ac{LMMSE} weight%
\footnote{Both \eqref{eqn:sigmasic} and \eqref{eqn:wlmmse} are approximations. The former neglects cross-user error correlations and inherits the known tendency of \ac{GaBP} to under-estimate posterior uncertainty, such that $\sigma_{\mathrm{tot}}^2$ is optimistic. The latter treats $\check{\bm{v}}_u$ as uncorrelated with the computing interference, which is justified in the regime $E_c \ll E_v$.}
follows as
\vspace{-1ex}
\begin{equation}
\vspace{-1ex}
\zeta_{\mathrm{lmmse}} = \frac{\eta U \sigma_c^2}{\eta^2 U \sigma_c^2 + \sigma_{\mathrm{tot}}^2},
\label{eqn:wlmmse}
\end{equation}
yielding the target function estimate $\hat{f} = \zeta_{\mathrm{lmmse}} r$.
The integrated receiver is summarized in Algorithm~1.

\vspace{-1ex}
\section{Performance Analysis}
\label{sec:results}
\vspace{-1ex}
To evaluate the proposed architecture, we consider a \ac{mu-MIMO} uplink over an uncorrelated block Rayleigh fading channel, with $M=4$ transmit antennas per user and computing symbols distributed as $c_u \sim \mathcal{CN}(0, 1)$.
The computing power is set as $E_c = 0.1\cdot E_v$, consistent with the $E_c \ll E_v$ regime, and the \ac{VGaBP} detector operates with a maximum of $\tau_{\text{max}}=30$ iterations and a damping factor of $\beta=0.5$.
Communication performance is measured via \ac{BER} over the $B$ bits of \eqref{eqn:rate} conveyed per user and channel use, while computing performance is evaluated via the \ac{NMSE} of the target function estimate
\vspace{-1ex}
\begin{equation}
\vspace{-1ex}
    \text{NMSE} \triangleq \frac{\mathbb{E}[|f-\hat{f}|^2]}{\mathbb{E}[|f|^2]},
\end{equation}
where $f = \sum_uc_u$ and $\mathbb{E}[|f|^2] = U\sigma_c^2$.
Both metrics are evaluated against $E_b/N_0$, where $E_b \triangleq (kE_v + E_c)/B$ denotes the total per-user radiated energy per bit.

\begin{figure}[t]
    \centering
    \includegraphics[width=1\columnwidth]{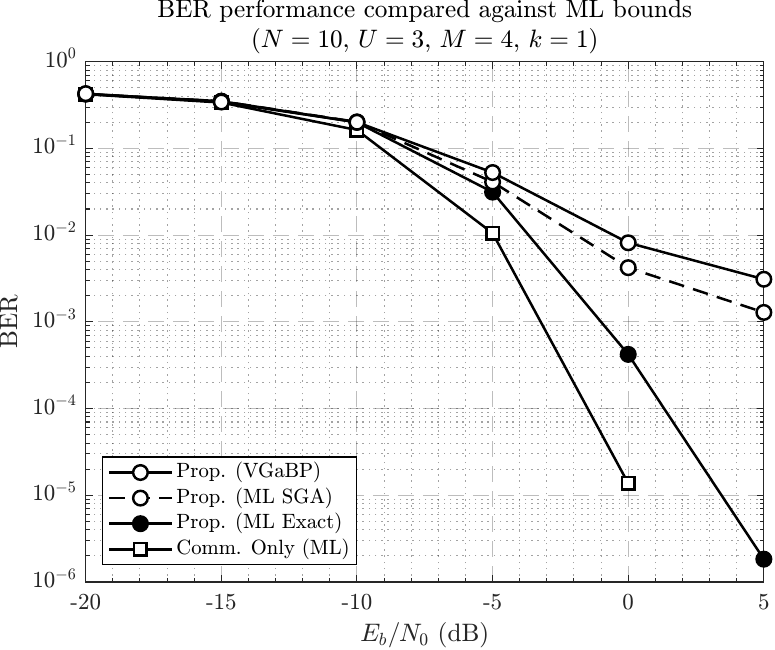}
    \caption{\ac{BER} performance compared with \ac{ML} baselines for a system with $N=10$, $M=4$, $U=3$ and $k=1$.}
    \label{fig:BER_ML}
    \vspace{-3ex}
\end{figure}

Since exact \ac{ML} detection scales as $\mathcal{O}(|\mathcal{V}|^U)$, we first benchmark a small-scale system with $N=10$, $U=3$ and $k=1$ under \ac{QPSK}, for which the search space remains tractable, against three \ac{ML} references.
The first disables the computing stream and redistributes the $E_c$ uniformly across the active data antennas, preserving the total radiated power while leaving the noise white.
The second retains the \ac{ICC} transmission and evaluates the true data-dependent interference covariance (\ac{ML} Exact), and the third replaces it by the averaged covariance of \eqref{eqn:Rexpected} (\ac{ML} \ac{SGA}).
As shown in Fig.~\ref{fig:BER_ML}, the \ac{VGaBP} detector tracks the \ac{SGA}-based \ac{ML} curve, confirming that the vector-valued message passing resolves the discrete codebook constraint at near-\ac{ML} accuracy under its own statistical model.
The remaining gap to the exact-covariance baseline therefore isolates the cost of the \ac{SGA}, which discards the correlation structure of the computing interference that the exact \ac{ML} receiver exploits.
The separation between exact-covariance curve and the communication-only reference in turn quantifies the intrinsic cost of embedding the computing stream.

\begin{figure}[t]
    \centering
    \includegraphics[width=1\columnwidth]{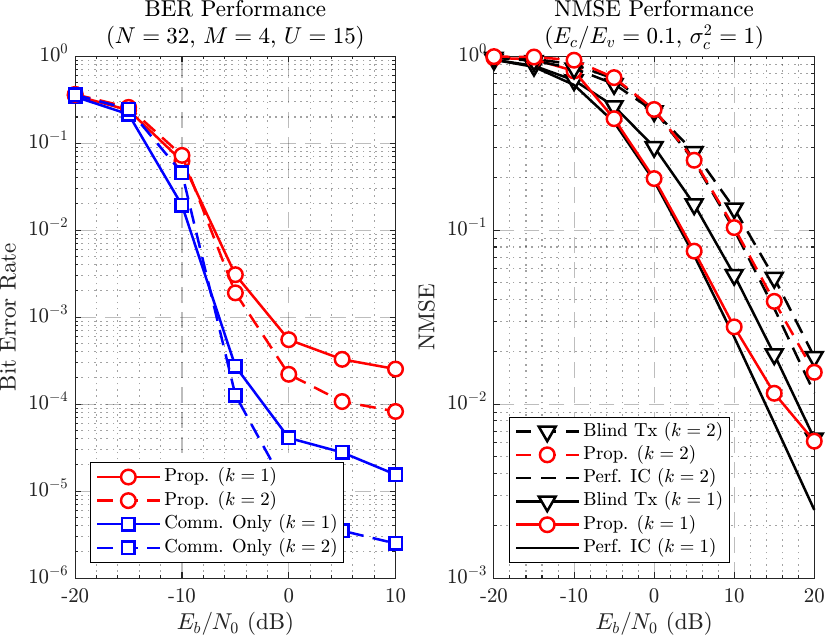}
    \caption{\ac{BER} (left) and \ac{NMSE} (right) performance for the proposed system with $N=32$, $M=4$, $U=15$ with $k=1$ and $k=2$ active antennas.}
    \label{fig:BER_NMSE}
    \vspace{-3ex}
\end{figure}

In Fig.~\ref{fig:BER_NMSE}, the system is scaled to $N=32$ and $U=15$, where \ac{ML} detection is computationally intractable and data detection is instead compared against a communication-only baseline employing the same \ac{VGaBP} algorithm.
For the \ac{AirComp} \ac{NMSE}, the proposed scheme is compared against two baselines.
The first is a channel-blind transmitter that spreads its computing power uniformly over the idle antennas without phase alignment, paired with a receiver-side vector \ac{LMMSE} combiner and perfect cancellation of the data term (Blind Tx).
The second is a variant of the proposed scheme in which the data term is likewise perfectly cancelled at the receiver (Perf. IC).
Since a blind transmitter faces no alignment constraint and would radiate its full budget $E_c$, it is constrained to the exact realized computing power of the proposed scheme, such that the comparison isolates the benefit of spatial alignment from the power actually radiated.
Under these identical power budgets, the blind baseline trails the proposed architecture, showing that the gain originates in the coherent alignment rather than in the radiated computing power.
Since \eqref{eqn:sic} subtracts the soft posterior mean $\bar{\bm{v}}_u$ rather than the hard estimate, the computing stream degrades with the residual uncertainty rather than with detection errors, such that the proposed scheme remains close to its perfect \ac{IC} bound.

We next analyze the spatial design trade-off by comparing $k=1$ with a \ac{QPSK} constellation against $k=2$ under \ac{BPSK}, thereby maintaining a constant rate of $B=4$ bits.
In the \ac{BER} evaluation, the $k=2$ configuration outperforms $k=1$, as spreading the same rate over two antennas reduces the required constellation order while leaving $|\mathcal{V}|$ , and therefore the detection complexity, unchanged.
The same reallocation, however, reduces the antennas available for \ac{AirComp} from $3$ to $2$, and the resulting loss of spatial \ac{DoF} increases the likelihood of deep fades that tighten the $\min_u \eta_u$ alignment bottleneck.
Allocating antennas to data transmission therefore improves communication performance at the direct expense of coherent alignment, which is the design trade-off of the proposed architecture.

Finally, we note that all \ac{BER} curves in Fig.~\ref{fig:BER_NMSE} exhibit an error floor at high \ac{SNR}, which is a documented convergence phenomenon of message passing algorithms settling on locally optimal fixed points in non-ideal high \ac{SNR} regimes \cite{Knoll_2022}.
The presence of such floors in the communication-only baseline, which carries no computing signal yet employs the same \ac{VGaBP} algorithm, confirms that they originate in the detector rather than in the \ac{AirComp} interference, which merely raises the floor level.
Addressing this effect through adaptively scaled belief updates \cite{Takahashi_2019} or other methods remains an orthogonal extension to this work.

\vspace{-1.5ex}
\section{Conclusion}
In this paper, we proposed a spatial \ac{ICC} architecture that exploits the inactive antennas of an \ac{IM} transmitter to integrate \ac{AirComp} capabilities.
By partitioning the array into active data antennas and pre-equalized computing antennas, the continuous computing waveform is isolated from the discrete digital payload at the transmitter, while at the receiver a \ac{VGaBP} detector recovers the data payload and a scalar \ac{LMMSE} filter estimates the target function from the residual.
Numerical results showed that the detector operates close to the \ac{ML} bound of its own statistical model, and that the number of antennas allocated to data transmission sets a direct trade-off between modulation robustness and \ac{AirComp} accuracy.

The residual gap to the exact-covariance \ac{ML} bound is attributable to the \ac{SGA} rather than to the message passing, so that closing it calls for detection under the true data-dependent covariance.
Since the cost of exact \ac{ML} grows exponentially with the network size, and the detection problem reduces to a search over a discrete unstructured candidate set, the \ac{ML} baselines established here can motivate quantum-accelerated search \cite{Rou_Asilomar2024} as a route to recovering that gain at scale.
\vspace{-1ex}

\section*{Acknowledgment}
This work was supported by the Deutsche Forschungsgemeinschaft (DFG, German Research Foundation) under the project QUBYSM, grant number $576171458$

\vspace{-1ex}
\bibliographystyle{IEEEtran}
\bibliography{references}

\end{document}